\documentclass[aps,pre,reprint,onecolumn,preprintnumbers,amsmath,amssymb,superscriptaddress,nofootinbib]{revtex4-2}
\pdfoutput=1

\usepackage{bm}
\usepackage{booktabs}
\usepackage{graphicx}
\usepackage{hyperref}
\usepackage{mathtools}
\usepackage{microtype}
\usepackage{array}

\hypersetup{
 colorlinks=true,
 linkcolor=blue,
 citecolor=blue,
 urlcolor=blue,
 pdftitle={Long-wavelength amplification of an NLS conserved-charge violation in a one-dimensional Gross-Pitaevskii-Poisson field},
 pdfauthor={Rin Takada},
 pdfkeywords={higher conserved charges, nonlinear Schrodinger equation, Gross-Pitaevskii-Poisson model, Jeans instability, long-range interactions, KPZ hydrodynamics}
}

\newcommand{\dd}{\mathrm{d}}
\newcommand{\ii}{\mathrm{i}}
\newcommand{\ee}{\mathrm{e}}
\newcommand{\order}{\mathcal{O}}
\newcommand{\mean}[1]{\left\langle #1\right\rangle}
\newcommand{\var}{\operatorname{Var}}
\newcommand{\cs}{c_{\mathrm{s}}}
\newcommand{\kJ}{k_{\mathrm{J}}}

\newcommand{\tinst}{t_{\mathrm{inst}}}
\newcommand{\Qthree}{Q_3}
\newcommand{\Fthree}{F_3}
\newcommand{\Pmom}{\mathcal{P}}
\newcommand{\Rfield}{R}

\begin{document}

\preprint{RESCEU-9/26}

\title{Long-wavelength amplification of an NLS conserved-charge violation in a one-dimensional Gross-Pitaevskii-Poisson field}

\author{Rin Takada}
\email{takada-rin@resceu.s.u-tokyo.ac.jp}
\affiliation{Research Center for the Early Universe (RESCEU), Graduate School of Science, The University of Tokyo, 7-3-1 Hongo, Bunkyo-ku, Tokyo 113-0033, Japan}

\date{July 23, 2026}

\begin{abstract}
Can weak self-gravity be treated as a small, generic perturbation of the integrable one-dimensional cubic nonlinear Schr\"odinger equation? We test this question by following the first nontrivial local NLS charge, $Q_3$. In the Gross-Pitaevskii-Poisson model it obeys the exact identity $\dot Q_3=\kappa F_3$, where $F_3$ is the instantaneous gravitational force on the charge. We quantify the initial violation through the normalized mean-square displacement of $Q_3$. Its curvature coefficient, $A_3=\operatorname{Var}(F_3)/\operatorname{Var}(Q_3)$, measures the root-mean-square initial speed of the charge in units of the charge's own ensemble width. For a UV-regulated, linearly stable Gaussian initial-data preparation, Wick contraction gives $\operatorname{Var}(Q_3)=\order(L)$ and $\operatorname{Var}(F_3)=\order(L^2)$, hence $A_3=\order(L)$. Direct Monte Carlo evaluation of the full nonlinear observables confirms these powers for $16\leqslant L\leqslant512$. The linear growth of $A_3$ means that admitting longer wavelengths makes nominally weak gravity increasingly effective at displacing the NLS charge. Along the Jeans-stable sequence $\kappa_L\propto L^{-2}$, the associated short-time curvature scale is $L^{3/2}$ instead of the $L^2$ benchmark of an infrared-uniform perturbation. This is an equal-time result, not a relaxation-rate or KPZ measurement. The local sound sector contains the Burgers self-coupling required for KPZ, but unscreened gravity also generates a relevant inverse-gradient coupling and, at fixed strength, a Jeans band. A dynamical connection to KPZ therefore requires a projected charge-relaxation calculation and a branch-resolved structure-factor test.
\end{abstract}

\maketitle

\begin{center}
\end{center}

\section{Introduction}
\label{sec:intro}

Generic one-dimensional nonintegrable fluids can display Kardar-Parisi-Zhang (KPZ) broadening of their sound peaks when the same-chirality sound self-coupling is nonzero~\cite{kardar1986,vanbeijeren2012,spohn2014}. Discrete Gross-Pitaevskii models provide a familiar realization, while integrable lattice models provide ballistic controls~\cite{kulkarni2015,mendl2015}. The continuum cubic nonlinear Schr\"odinger equation (NLS) sits on the integrable side of this divide: it possesses infinitely many local conserved charges~\cite{zakharov1972,zakharov1973,faddeev2007,adams2024}, and this hierarchy obstructs the generic hydrodynamic closure on which the KPZ prediction rests. A phonon linewidth close to $|k|^{3/2}$ therefore does not, by itself, show that the continuum NLS has become a generic stochastic KPZ fluid~\cite{kulkarni2013,bouchoule2023}. For the continuum equation, any KPZ scenario must begin with a perturbation that destroys the extra charges.

Self-gravity is a natural candidate for that perturbation: it is not compatible with the NLS hierarchy, and it couples widely separated wavelengths. But it is not a generic perturbation. Its Poisson kernel is proportional to $1/k^2$, so its strength is concentrated at the longest wavelengths, and at fixed coupling it produces the Jeans instability in precisely the infrared sector in which hydrodynamics is normally sought. Calling such a perturbation ``weak'' is therefore ambiguous until one specifies at which wavelength its strength is measured. This fixes the central question of the paper: does the charge breaking caused by weak self-gravity remain uniformly weak as progressively longer wavelengths are admitted?

We answer that question with a concrete charge diagnostic. The first nontrivial local NLS charge $Q_3$ obeys an exact Gross-Pitaevskii-Poisson (GPP) identity, $\dot Q_3=\kappa F_3$, so the functional $F_3$ is the instantaneous gravitational force on the charge. To decide whether this force is large or small in a given ensemble, we measure it against the only intrinsic reference scale available: the ensemble width of $Q_3$ itself. The resulting coefficient $A_3=\var(F_3)/\var(Q_3)$ is the squared initial speed of the charge in units of the charge's own fluctuations, and it controls the short-time curvature of the normalized mean-square displacement of $Q_3$. Both a Wick calculation and a direct Monte Carlo evaluation give $A_3\propto L$ for the periodic length $L$. Since $L$ sets the infrared cutoff, this linear growth is a direct demonstration that the charge-breaking effect of weak self-gravity is not uniform in the infrared.

The result is deliberately an equal-time statement. Existing work on weak integrability breaking and generalized hydrodynamics primarily addresses projected long-time collision operators or hydrodynamic crossovers~\cite{doyonyoshimura2017,durnin2021,hutsalyuk2021,friedman2020,lopez2021,bastianello2021,panfil2023,lebek2024,lebek2025,biagetti2026,jung2007}; those descriptions require a stationary reference state and a zero-frequency force spectrum, neither of which is computed here. What is new here is complementary to that program: an exact, model-specific force law and the infrared scaling of its equal-time strength.

For KPZ the result yields a constraint rather than an observation. The local sound equations of the GPP model do contain the Burgers self-coupling that KPZ broadening requires, but an actual KPZ regime also requires decay of the extra integrable charges and a stochastic nonlinear-fluctuating-hydrodynamic closure, and the present calculation establishes neither. What it does show is that unscreened gravity cannot be treated as a smooth, size-independent local perturbation: the same infrared nonuniformity that drives $A_3\propto L$ reappears at the hydrodynamic level as an enhanced short-time charge response, a relevant inverse-gradient sound coupling, and a Jeans band at fixed nonzero strength.

The paper proceeds as follows. Section~\ref{sec:model} defines the model and the infrared limits in which the question is posed. Section~\ref{sec:charge} derives the exact force law and the normalized diagnostic. Section~\ref{sec:infrared} establishes $A_3\propto L$ analytically and numerically. Section~\ref{sec:kpz} translates the result into the two measurements that a KPZ claim would still require, and Sec.~\ref{sec:conclusion} concludes.

\section{Model and the infrared question}
\label{sec:model}

We work on a periodic interval $x\in[0,L)$. We call it a ``box,'' but it is only an infrared regulator used to define Fourier modes, not a literal astronomical container. Its smallest nonzero wave number is $k_1=2\pi/L$. Increasing $L$ therefore admits longer wavelengths and probes the infrared limit.

With $\hbar=m=1$ and $g,\kappa>0$, the one-dimensional GPP equations are
\begin{align}
 \ii\partial_t\psi
 &= -\frac{1}{2}\partial_x^2\psi+g|\psi|^2\psi+\Phi\psi,
 \label{eq:gpp}\\
 \partial_x^2\Phi
 &=\kappa(\rho-\bar\rho),
 \qquad \rho=|\psi|^2,
 \qquad \bar\rho=L^{-1}\int_0^L\rho\,\dd x .
 \label{eq:poisson}
\end{align}
The zero mode of $\Phi$ is set to zero, so $\Phi_k=-\kappa\rho_k/k^2$ for $k\ne0$. With $\Phi$ understood as this zero-mean Poisson solution, the reduced Hamiltonian is
\begin{equation}
 H=\int_0^L\dd x\left[
 \frac{1}{2}|\partial_x\psi|^2+\frac{g}{2}\rho^2
 +\frac{1}{2}(\rho-\bar\rho)\Phi\right].
 \label{eq:H}
\end{equation}
Its gravitational part is nonpositive, while particle number, momentum $\Pmom=\int\rho u\,\dd x$, and $H$ are conserved.

Writing $\psi=\sqrt\rho\,\ee^{\ii\theta}$ and $u=\partial_x\theta$ gives, wherever $\rho>0$,
\begin{align}
 &\partial_t\rho+\partial_x(\rho u)=0,
 \label{eq:continuity}\\
 &\partial_tu+u\partial_xu+g\partial_x\rho+\partial_x\Phi
 +\partial_xQ[\rho]=0,
 \qquad Q[\rho]=-\frac{1}{2}\frac{\partial_x^2\sqrt\rho}{\sqrt\rho}.
 \label{eq:euler}
\end{align}
Linearization about $\rho=\rho_0$ yields
\begin{equation}
 \omega_{\rm sg}^2(k)=\cs^2k^2+\frac{k^4}{4}-\rho_0\kappa,
 \qquad \cs^2=g\rho_0,
 \label{eq:dispersion}
\end{equation}
and the Jeans wave number
\begin{equation}
 \kJ^2=2\left(\sqrt{\cs^4+\rho_0\kappa}-\cs^2\right).
 \label{eq:kJ}
\end{equation}
The stable sound sector has $|k|>\kJ$; modes with $0<|k|<\kJ$ grow exponentially. Jeans spectra in GPP systems are well known~\cite{chavanis2020}. Here the spectrum is used only to distinguish the stable and unstable infrared limits relevant to the charge-breaking analysis.

The infrared question involves two limits that do not commute, and two dimensionless parameters keep them separate:
\begin{equation}
 \mu=\frac{\rho_0\kappa}{\cs^4}=\frac{\kappa}{g^2\rho_0},
 \qquad
 \lambda=\frac{L\kJ}{2\pi}=\frac{\kJ}{k_1}.
 \label{eq:mu-lambda}
\end{equation}
The parameter $\mu$ measures the gravitational coupling against the self-interaction, while $\lambda$ counts the number of linearly unstable Fourier modes; for $\mu\ll1$, $\kJ/\cs=\mu^{1/2}(1-\mu/8+\cdots)$. At fixed $L$, taking $\mu\to0$ eventually removes all unstable Fourier modes. At fixed $\kappa>0$, taking $L\to\infty$ inevitably introduces a Jeans band. Holding $\lambda$ fixed below unity therefore defines the only linearly stable route to the infrared: a double scaling in which $\kappa$ decreases as $L$ grows. The value $\lambda=1$ is marginal, and $\lambda>1$ leaves the lowest mode unstable.

The unstable alternative sets a hard time limit. For unfiltered initial conditions with an unstable lowest mode,
\begin{equation}
 \gamma_1=\left(\rho_0\kappa-\cs^2k_1^2-\frac{k_1^4}{4}\right)^{1/2},
 \qquad
 \tinst\simeq\gamma_1^{-1}\ln\!\left(\frac{c\rho_0}{\sigma_{\rm J}}\right),
 \label{eq:tinst}
\end{equation}
where $\sigma_{\rm J}$ is the initial root-mean-square amplitude in the unstable band. Any homogeneous hydrodynamic window must lie before this background-breakdown scale. The stable double-scaling sequence used below is chosen to avoid this breakdown entirely. Its purpose is not to describe a particular physical system at fixed $\kappa$; it is a controlled way of asking whether weak gravity remains uniform as the infrared cutoff is removed. The scaling resembles a Kac-type size adjustment in long-range systems, but here it is fixed by Jeans stability rather than by a standard extensivity prescription~\cite{campa2009}.

\section{What gravity does to an NLS charge}
\label{sec:charge}

\subsection{Why follow a higher charge?}

Integrability is stronger than conservation of particle number, momentum, and energy. The cubic NLS has an infinite hierarchy of additional local conserved functionals. The first nontrivial one in the normalization of Eq.~\eqref{eq:gpp} is~\cite{faddeev2007,adams2024}
\begin{equation}
 \Qthree=\operatorname{Re}\left\{\ii\int_0^L\dd x\left[
 (\partial_x\psi)(\partial_x^2\psi^*)
 +3g|\psi|^2\psi\,\partial_x\psi^*\right]\right\}.
 \label{eq:Q3}
\end{equation}
For $\kappa=0$, $Q_3$ is exactly conserved. A nonzero $\dot Q_3$ therefore gives a direct diagnostic of the loss of the NLS charge hierarchy. It is more specific than an energy-exchange argument and should not be confused with a proof of thermalization.

For the full GPP dynamics, differentiation of Eq.~\eqref{eq:Q3} and periodic integration by parts give the exact identity
\begin{equation}
 \dot\Qthree=\kappa\Fthree,
 \qquad
 \Fthree=\int_0^L\dd x\,\frac{\partial_x\Phi}{\kappa}
 \left[
 \partial_x^2\rho-\frac{3}{4}\frac{(\partial_x\rho)^2}{\rho}
 -3\rho u^2-3g\rho^2
 \right].
 \label{eq:F3}
\end{equation}
Thus $F_3$ is the instantaneous gravitational force on the charge: for a given field configuration it specifies the initial direction and speed with which gravity moves $Q_3$. Equation~\eqref{eq:F3} is exact wherever $\rho>0$.

\subsection{What is normalized, and why?}

The raw value of $Q_3$ grows with system size, so a change in $Q_3$ cannot be compared across different $L$ without a reference scale. We use the natural root-mean-square width of $Q_3$ in the chosen initial ensemble and define
\begin{equation}
 \mathcal D_3(t)=
 \frac{\mean{[\Qthree(t)-\Qthree(0)]^2}}{\var(\Qthree)}.
 \label{eq:D3-definition}
\end{equation}
A value $\mathcal D_3=1$ means that the root-mean-square displacement of the charge is as large as a typical fluctuation of the charge itself. Since $\dot Q_3=\kappa F_3$, the time-reversal-symmetric preparations used below give
\begin{equation}
 \mathcal D_3(t)=\kappa^2A_3(L)t^2+\order(t^4),
 \qquad
 A_3(L)=\frac{\var(\Fthree)}{\var(\Qthree)}.
 \label{eq:A3-definition}
\end{equation}
This equation gives $A_3$ a direct meaning. The quantity $\kappa\sqrt{A_3}$ is the root-mean-square initial speed of $Q_3$, measured in units of the root-mean-square width of $Q_3$. At fixed $\kappa$, a larger $A_3$ means that the normalized charge departs more rapidly from its conserved value at short times. The corresponding curvature scale is
\begin{equation}
 \tau_{3,\rm curv}=\frac{1}{\kappa\sqrt{A_3}}.
 \label{eq:curvature-def}
\end{equation}
This is a short-time Taylor scale, not a decay time or a kinetic relaxation rate.

\section{Why long wavelengths amplify the charge violation}
\label{sec:infrared}

\subsection{A stable sequence and an initial-data ensemble}

To approach longer wavelengths without triggering the lowest Jeans mode, we choose
\begin{equation}
 \kappa_L=\eta gk_1^2,
 \qquad 0\leqslant\eta<1.
 \label{eq:kappa-eta}
\end{equation}
By Eqs.~\eqref{eq:kJ} and \eqref{eq:mu-lambda}, this choice gives $\lambda^2\leqslant\eta<1$ at every $L$, with $\lambda\to\sqrt\eta$ as $L\to\infty$, so it realizes the linearly stable double scaling of Sec.~\ref{sec:model} with $\eta$ as its single control parameter. The initial fields are drawn from the UV-regulated quadratic distribution
\begin{equation}
 \mathcal P_\eta[\delta\rho,u]\propto\exp[-H_{2,\eta}/T],
 \quad
 H_{2,\eta}=\frac{L}{2}\sum_{0<|k|\leqslant k_{\rm UV}}
 \left\{\rho_0|u_k|^2+
 \left[g+\frac{k^2}{4\rho_0}-\frac{\kappa_L}{k^2}\right]|\delta\rho_k|^2\right\}.
 \label{eq:phonon-ensemble}
\end{equation}
This is a Gaussian initial-data preparation defined by the linearized Hamiltonian, not the full nonlinear GPP Gibbs state. It is well defined along the entire sequence, because the density stiffness $g+k^2/4\rho_0-\kappa_L/k^2$ in Eq.~\eqref{eq:phonon-ensemble} is positive for every retained mode when $\eta<1$; and because the distribution is even in $u$, it is time-reversal symmetric, which is the property that removes the linear-in-$t$ term and produces the pure $t^2$ curvature in Eq.~\eqref{eq:A3-definition}. For evaluation of the full nonlinear observables, the numerical sampling is additionally conditioned on positive density at every grid point. No primary production draw was rejected; the smallest observed density was $0.3723$ in the dressed preparation and $0.3134$ in the bare control.

At leading weak-fluctuation order,
\begin{align}
 \Qthree^{(2)}
 &=\int_0^L\dd x\,[6g\rho_0\delta\rho\,u+(\partial_x\delta\rho)(\partial_xu)],
 \label{eq:Q3-leading}\\
 \Fthree^{(3)}
 &=-3\int_0^L\dd x\,\Rfield
 \left[\rho_0u^2+g(\delta\rho)^2+
 \frac{(\partial_x\delta\rho)^2}{4\rho_0}\right],
 \qquad \Rfield=\partial_x^{-1}\delta\rho.
 \label{eq:F3-leading}
\end{align}

\subsection{The scaling argument}

The two variances scale differently for a transparent reason. The charge $Q_3$ is the integral of a local density. With a fixed UV cutoff, ordinary extensivity gives
\begin{equation}
 \var(Q_3)=\order(L).
 \label{eq:varQ}
\end{equation}
The force $F_3$ contains the inverse derivative $R=\partial_x^{-1}\delta\rho$. In Fourier space, $R_k=\delta\rho_k/(\ii k)$, so the longest available mode is amplified by $1/k_1\propto L$. Along the stable sequence this gives $\langle R^2\rangle=\order(L)$. The spatial integral of the resulting force density then has
\begin{equation}
 \var(F_3)=\order(L^2).
 \label{eq:varF}
\end{equation}
Appendix~\ref{app:charge} gives an explicit positive Wick contraction and shows that the $L^2$ term cannot cancel. Consequently,
\begin{equation}
 A_3(L)=\order(L).
 \label{eq:A3-prediction}
\end{equation}
Thus the normalized short-time curvature grows linearly with the
infrared length scale. It means that the gravitational force on the NLS charge is not controlled by a size-independent weak-coupling coefficient. Allowing longer wavelengths increases the normalized short-time action of gravity.

Along the stable sequence the response still slows with increasing
system size, but more slowly than for an infrared-uniform
perturbation. At fixed $\kappa$, the normalized initial speed $\kappa\sqrt{A_3}$ grows as $L^{1/2}$. Along the stable sequence, however, $\kappa_L\propto L^{-2}$, so the absolute response still becomes slower:
\begin{equation}
 \kappa_L^2A_3(L)=\order(L^{-3}),
 \qquad
 \tau_{3,\rm curv}=\order(L^{3/2}).
 \label{eq:curvature-time}
\end{equation}
Without the infrared enhancement, an $L$-independent $A_3$ would instead give $\tau_{3,\rm curv}=\order(L^2)$. The change from $L^2$ to $L^{3/2}$ is therefore the operational consequence of the long-wavelength amplification.

\subsection{Monte Carlo check}

We evaluated the full nonlinear $Q_3$ and $F_3$ without time integration. The primary gravity-dressed preparation used $\eta=1/2$ and $3000$ independent realizations per length; the bare-NLS control used $\eta=0$ and $5000$. Both used $g=\rho_0=1$, $T=0.02$, $k_{\rm UV}=3$, $\Delta x=0.25$, and eleven lengths $16\leqslant L\leqslant512$. Pure-power fits over $L\geqslant64$ give
\begin{align}
 \text{dressed:}\quad
 p_Q&=1.025\ [0.998,1.054],&
 p_F&=2.012\ [1.973,2.053],&
 p_A&=0.987\ [0.937,1.033],
 \label{eq:dressed-fits}\\
 \text{bare:}\quad
 p_Q&=1.020\ [0.998,1.042],&
 p_F&=2.018\ [1.987,2.045],&
 p_A&=0.998\ [0.961,1.033].
 \label{eq:bare-fits}
\end{align}
The brackets are pair-bootstrap $95\%$ intervals for the stated fit model and window. Theil-Sen, Huber, leave-one-out, fit-window, and $1/L$-correction analyses are compatible with $(p_Q,p_F,p_A)=(1,2,1)$. Additional dressed suites varying $T$, $k_{\rm UV}$, and $\eta$ give finite-window $p_A$ values from $1.002$ to $1.065$. The sampled $F_3$ distributions are leptokurtic, so pair bootstrapping is used for the pointwise error bars.

\begin{figure}[t]
 \centering
 \includegraphics[width=0.96\linewidth]{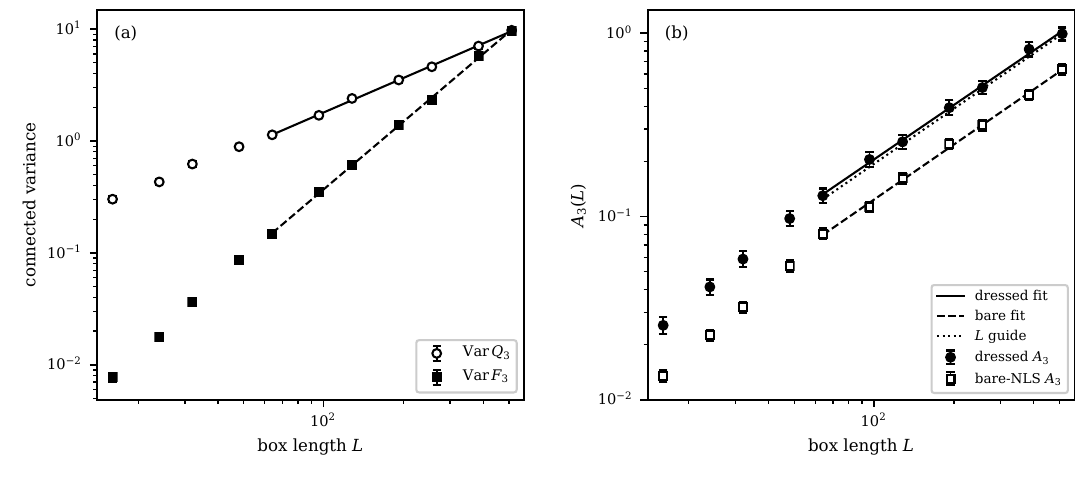}
 \caption{Finite-size scaling of the higher-charge observables. (a) Connected variances in the gravity-dressed stable preparation. (b) The normalized short-time curvature coefficient $A_3$ in dressed and bare-NLS preparations. Fits use only $L\geqslant64$ and are drawn only over that range; the dotted line is an $L$ guide. Error bars are pointwise pair-bootstrap $95\%$ intervals.}
 \label{fig:higher-charge}
\end{figure}

The normalization is not unique under addition of lower conserved charges. Since momentum $\Pmom$ remains conserved, $Q_3+c\Pmom$ has the same force but a different variance. We use the standard hierarchy charge. As a check, the alternative $Q_3-6g\rho_0\Pmom$ gives $A_3\propto L^{0.999}$ in the dressed data and $L^{1.005}$ in the bare data. The exponent is robust, whereas the prefactor and the absolute curvature time are charge-basis dependent.

\subsection{Equal-time curvature and relaxation rates}
\label{sec:A3-scope}

The equal-time coefficient $A_3$ must not be mistaken for a relaxation rate, and it is worth stating exactly which information it lacks. In a stationary reference state, with all slow charges projected out, a memory-matrix description has the schematic form~\cite{jung2007,durnin2021,bastianello2021}
\begin{equation}
 M_{ab}=\int_0^\infty\dd t\,
 \mean{\widehat F_a(t)\widehat F_b(0)}_{\rm stat,c},
 \qquad \bm\Gamma=\kappa^2\bm M\bm\chi^{-1}.
 \label{eq:memory-matrix}
\end{equation}
The present initial-data preparation is not shown to be stationary under the nonlinear dynamics, and no projected zero-frequency force spectrum is computed here. The result $A_3\propto L$ is therefore an equal-time statement, not a long-time relaxation law.

\section{How this result connects to KPZ}
\label{sec:kpz}

The route from the present charge result to KPZ contains two separate dynamical questions. First, do the extra NLS charges actually relax so that a conventional hydrodynamic description becomes valid? Second, if they do, do the sound peaks enter the KPZ universality class? The calculation above addresses the initial violation of one charge and the infrared nonuniformity of that violation; it does not answer either long-time question. We first identify the local Burgers coupling, then examine the
infrared gravitational term and the dynamical tests required for KPZ. Sec.~\ref{sec:kpz-ingredient} identifies the one KPZ ingredient the model already contains, Sec.~\ref{sec:kpz-obstruction} explains why unscreened gravity nevertheless obstructs the naive infrared limit, and Sec.~\ref{sec:kpz-tests} states the two measurements that would settle the dynamical questions.

\subsection{The local KPZ ingredient}
\label{sec:kpz-ingredient}

In the local sound regime, define
\begin{equation}
 \varphi_\sigma=\frac{1}{2}\left(u+\sigma\frac{\cs}{\rho_0}\delta\rho\right),
 \qquad \sigma=\pm.
 \label{eq:local-branches}
\end{equation}
To quadratic Euler order,
\begin{equation}
 (\partial_t+\sigma\cs\partial_x)\varphi_\sigma
 =-\frac{3}{2}\varphi_\sigma\partial_x\varphi_\sigma
 +\mathcal C_\sigma[\varphi_+,\varphi_-]
 -a_g\partial_x^{-1}(\varphi_+-\varphi_-)+\mathcal D_\sigma,
 \qquad a_g=\frac{\rho_0\kappa}{2\cs}.
 \label{eq:branch-schematic}
\end{equation}
The explicit quadratic coefficients contained in $\mathcal C_\sigma$, together with the projection of the gravitational term and the value of $a_g$, are derived in Appendix~\ref{app:branches}. The nonzero coefficient of $\varphi_\sigma\partial_x\varphi_\sigma$ is the deterministic Burgers ingredient behind KPZ sound broadening in generic one-dimensional fluids. It is necessary but not sufficient. A KPZ claim also needs a stationary state, decay of the extra integrable charges, the full susceptibility structure including the energy mode, dissipation, noise, and a scale separation that justifies nonlinear fluctuating hydrodynamics.

\subsection{Why unscreened gravity changes the infrared limit}
\label{sec:kpz-obstruction}

The infrared nonuniformity established for the charge in Sec.~\ref{sec:infrared} is mirrored by three related hydrodynamic manifestations of the same $1/k^2$ Poisson kernel. They are not independent microscopic mechanisms: they concern, respectively, the scaling of the projected sound equation, the nonanalytic correction to the stable dispersion, and the unstable Jeans sector of the fixed-$\kappa$ thermodynamic limit. First, at a local KPZ fixed point, $x\to bx$, $t\to b^{3/2}t$, and the slope scales as $\varphi\to b^{-1/2}\varphi$. The inverse-gradient coefficient therefore flows as
\begin{equation}
 a_g\to b^{5/2}a_g,
 \label{eq:ag-flow}
\end{equation}
so the unscreened gravitational term is tree-level relevant. Second, the same nonuniformity appears in the stable dispersion for $\kJ\ll|k|\ll\cs$,
\begin{equation}
 \omega_{\rm sg}(k)=\cs|k|+\frac{|k|^3}{8\cs}
 -\frac{\rho_0\kappa}{2\cs|k|}+\cdots.
 \label{eq:dispersion-expand}
\end{equation}
Third, at fixed nonzero $\kappa$, the homogeneous thermodynamic limit contains a Jeans band. These are mathematically distinct diagnostics of one common infrared singularity. Taken together, they show that one cannot obtain an asymptotic homogeneous local-KPZ theory by simply declaring weak unscreened gravity to be a small local integrability-breaking perturbation.

This does not exclude a finite-time or infrared-filtered KPZ-like window. Such a window would have to satisfy three conditions simultaneously: the extra NLS charges must no longer control the dynamics, the background must remain usable, and the inverse-gradient correction must be subleading over the measured wave-number range. The linearly growing Jeans sector cannot simply be assumed to provide white noise; its covariance is nonstationary, and two unstable modes directly force only $|k|<2\kJ$.

\subsection{The two calculations still needed}
\label{sec:kpz-tests}

A controlled connection to KPZ requires one measurement per question posed at the start of this section. For the first question, the projected force spectrum in Eq.~\eqref{eq:memory-matrix} must determine whether and when the higher-charge hierarchy relaxes. For the second, the branch-resolved dynamic structure factor must be measured after recentering by the exact gravitational dispersion. A local KPZ sound peak would obey
\begin{equation}
 S_{\sigma\sigma}(x,t)\simeq
 \frac{1}{(\lambda_s t)^{2/3}}
 f_{\rm KPZ}\!\left(
 \frac{x-\sigma\cs t}{(\lambda_s t)^{2/3}}
 \right)
 \label{eq:kpz-test}
\end{equation}
within a common time and wave-number window. Comparing unscreened gravity with a local quintic perturbation and a screened Poisson kernel would separate generic integrability breaking from the specific $1/k^2$ infrared singularity. A quintic control remains locally stable when $g\rho_0+2\epsilon\rho_0^2>0$, while the screened model
\begin{equation}
 (\partial_x^2-m_s^2)\Phi=\kappa(\rho-\bar\rho)
 \label{eq:screened}
\end{equation}
is acoustically stable for $\kappa<gm_s^2$. Recent Bose-gas studies provide complementary phonon-relaxation observables, but they do not replace these two model-specific tests~\cite{alilou2026}.

\section{Conclusion}
\label{sec:conclusion}

We derived the exact force law $\dot Q_3=\kappa F_3$ and found that
its normalized equal-time variance grows linearly with system size. The primary result is the exact force law $\dot Q_3=\kappa F_3$ together with the infrared scaling $A_3=\var(F_3)/\var(Q_3)\propto L$. In words, the longest wavelengths amplify the short-time gravitational displacement of an NLS conserved charge relative to the natural width of that charge. Along a linearly stable $\kappa_L\propto L^{-2}$ sequence, this changes the curvature scale from the infrared-uniform benchmark $L^2$ to $L^{3/2}$. Weak unscreened gravity is therefore not a size-independent local perturbation of the integrable NLS.

The implication for KPZ is a constraint, not an observation. The sound equation contains the required Burgers self-coupling, but unscreened gravity also introduces a relevant inverse-gradient term and a Jeans band. Establishing a transient KPZ regime requires a projected long-time charge-relaxation calculation and an exact-dispersion-centered structure-factor measurement.

\begin{acknowledgments}
The author is grateful to K. Hotokezaka, R. Jinno, and R. Namba for helpful comments that improved the manuscript.
\end{acknowledgments}

\section*{Data availability}

The numerical data that support the findings of this study are
available from the author upon reasonable request. The source code
is not publicly available.

\appendix

\section{Higher-charge identity and Wick scaling}
\label{app:charge}

Under the potential flow, $\partial_t\psi=-\ii\Phi\psi$. Differentiating Eq.~\eqref{eq:Q3} gives
\begin{align}
 \partial_t\Qthree
 =\operatorname{Re}\int_0^L\dd x\,[
 &\Phi_x\psi\psi_{xx}^* -\Phi_{xx}\psi_x\psi^*
 -2\Phi_x|\psi_x|^2-3g\Phi_x|\psi|^4].
 \label{eq:Q3-intermediate}
\end{align}
One integration by parts and
\begin{equation}
 2\operatorname{Re}(\psi^*\psi_{xx})-|\psi_x|^2
 =\rho_{xx}-\frac{3}{4}\frac{\rho_x^2}{\rho}-3\rho u^2
 \label{eq:hydro-identity}
\end{equation}
produce Eq.~\eqref{eq:F3}.

For the Fourier convention $f(x)=\sum_k f_k\ee^{\ii kx}$,
\begin{equation}
 \Qthree^{(2)}=L\sum_{k\ne0}(6g\rho_0+k^2)\delta\rho_k u_{-k}.
 \label{eq:Q3-fourier}
\end{equation}
Using the mode variances from Eq.~\eqref{eq:phonon-ensemble}, the number of retained modes is $\order(L)$ and hence $\var(Q_3)=\order(L)$.

Let $\Rfield=\partial_x^{-1}\delta\rho$, $C_R(r)=\mean{\Rfield(x)\Rfield(x+r)}$, and $C_u(r)=\mean{u(x)u(x+r)}$. The velocity part of Eq.~\eqref{eq:F3-leading} is
\begin{equation}
 F_u=-3\rho_0\int_0^L\dd x\,\Rfield(x)[u(x)^2-C_u(0)],
 \label{eq:Fu}
\end{equation}
because $\int\Rfield\,\dd x=0$. Density and velocity are independent in Eq.~\eqref{eq:phonon-ensemble}, so the covariance of $F_u$ with the density-only part of $\Fthree^{(3)}$ vanishes. Wick's theorem gives the nonnegative contribution
\begin{align}
 \var(F_u)
 &=18\rho_0^2L\int_0^L\dd r\,C_R(r)C_u(r)^2
 \nonumber\\
 &=18\rho_0^2L^2\sum_{k,q}S_R(k)S_u(q)S_u(-k-q),
 \label{eq:Fu-variance}
\end{align}
where $S_R$ and $S_u$ are nonnegative Fourier spectra. The lowest mode has $S_R(k_1)=\order(L)$, while $\sum_qS_u(q)S_u(-k_1-q)=\order(L^{-1})$ at fixed cutoff. Thus Eq.~\eqref{eq:Fu-variance} contains a strictly positive $\order(L^2)$ term. The remaining density-only variance is nonnegative, so the leading power cannot cancel.

More explicitly,
\begin{equation}
 C_R(0)=\frac{TL}{2\pi^2g}
 \sum_{n=1}^{n_{\rm UV}}\frac{1}{n^2-\eta+\order(n^4/L^2)}
 =L\,C(\eta)+\order(1),
 \label{eq:primitive-variance}
\end{equation}
with finite $C(\eta)$ for fixed $\eta<1$ and $C(\eta)\sim T/[2\pi^2g(1-\eta)]$ only near $\eta\to1^-$. At fixed hard cutoff, the real-space correlations are integrable cutoff correlations rather than strictly finite-range functions.

\section{Quadratic branch coefficients}
\label{app:branches}

This appendix gives the explicit projection underlying Eq.~\eqref{eq:branch-schematic} and the same-chirality Burgers coefficient used in Sec.~\ref{sec:kpz-ingredient}. With $u=\varphi_++\varphi_-$ and $\delta\rho=(\rho_0/\cs)(\varphi_+-\varphi_-)$, the local Euler terms give
\begin{align}
 (\partial_t+\cs\partial_x)\varphi_+
 =&-\frac{3}{2}\varphi_+\partial_x\varphi_+
 -\frac{1}{2}\varphi_+\partial_x\varphi_-
 -\frac{1}{2}\varphi_-\partial_x\varphi_+
 +\frac{1}{2}\varphi_-\partial_x\varphi_-
 \nonumber\\
 &-a_g\partial_x^{-1}(\varphi_+-\varphi_-)+\mathcal D_+,
 \label{eq:phi-plus-app}\\
 (\partial_t-\cs\partial_x)\varphi_-
 =&-\frac{3}{2}\varphi_-\partial_x\varphi_-
 -\frac{1}{2}\varphi_+\partial_x\varphi_-
 -\frac{1}{2}\varphi_-\partial_x\varphi_+
 +\frac{1}{2}\varphi_+\partial_x\varphi_+
 \nonumber\\
 &-a_g\partial_x^{-1}(\varphi_+-\varphi_-)+\mathcal D_-.
 \label{eq:phi-minus-app}
\end{align}
The coefficient $a_g=\rho_0\kappa/(2\cs)$ follows from $-\partial_x\Phi/2$ in either projected equation.

\nocite{bastianello2018,koch2022,levkov2018,mukherji1997,katzav2003}
\bibliographystyle{apsrev4-2}
\bibliography{main}

\end{document}